\documentclass[usenatbib,fleqn]{mnras}

\usepackage{amsmath} 
\usepackage{amssymb}	

\title{Vortex buoyancy in superfluid and superconducting neutron stars}

\author[V.~A.~Dommes, M.~E.~Gusakov]
{V.~A.~Dommes,
 M.~E.~Gusakov
 \\
Ioffe Institute,
Polytekhnicheskaya 26, 194021 St.~Petersburg, Russia
}

\date{}

\renewcommand{\v}[1]{\ensuremath{\mbox{\boldmath$ #1 $}}}  
\newcommand{\gv}[1]{\ensuremath{\mbox{\boldmath$ #1 $}}}  
\newcommand{\pd}[2]{\frac{\upartial #1}{\upartial #2}} 
 
\newcommand{\rot}[1]{\operatorname{curl} #1} 

\def\pFn{p_{\raise-0.3ex\hbox{{\scriptsize F$\!$\raise-0.03ex\hbox{\rm n}}}}
}  
\def\pFp{p_{\raise-0.3ex\hbox{{\scriptsize F$\!$\raise-0.03ex\hbox{\rm p}}}}
}  
\def\pFe{p_{\raise-0.3ex\hbox{{\scriptsize F$\!$\raise-0.03ex\hbox{\rm e}}}}
}  
\def\pFmu{p_{\raise-0.3ex\hbox{{\scriptsize F$\!$\raise-0.03ex\hbox{\rm
					$\mu$}}}} }  

\begin{document}

\maketitle

\begin{abstract}
Buoyancy of proton vortices is considered as one of the important mechanisms
of magnetic field expulsion from the superconducting interiors of neutron stars.
Here we show
that the generally accepted expression for the buoyancy force
is not correct and should be modified.
The correct expression is derived for both neutron and proton vortices.
It is argued that this force
is already contained in the coarse-grained hydrodynamics of Bekarevich \& Khalatnikov
and its various multifluid extensions, but is absent in the hydrodynamics of Hall. 
Some potentially interesting buoyancy-related effects are briefly discussed.
\end{abstract}

\begin{keywords}
stars: neutron --  stars: magnetic fields --  stars: interiors -- hydrodynamics
\end{keywords}

\section{Introduction}

According to theoretical calculations
(e.g., \citealt{ls01,plps13}), 
protons in the neutron star cores
condense into a superconducting state
at temperatures below
$T \sim 10^8 - 10^{10}~{\rm K}$.
If they form a type-II superconductor,
the magnetic flux penetrates the core
in the form of quantized fluxtubes 
(also called Abrikosov vortices 
or, simply, proton vortices),
each carrying a single flux quantum
$\phi_0 \equiv \pi \hbar c / e_{\rm p}$,
where $\hbar$, $c$, and $e_{\rm p}$ 
are the Planck constant, speed of light, and proton electric charge, 
respectively.

Since the magnetic field in the neutron star core is locked to proton vortices,
the problem of its evolution reduces to that of vortex motion.
This motion is, in turn, determined by the balance of forces acting on vortices.
One of these forces is the so called
{\it buoyancy force},
introduced into the neutron-star literature by \citet{1985Ap&SS.115...43M}
and used subsequently in many works studying the magnetic field expulsion 
from the superconducting interiors of neutron stars 
\citep[e.g.][]{hrs86,jones87, sbmt90, ccd92, dcc93,chau97, miri00,miri02, kg00, kg01a,
2016MNRAS.456.4461E}.
Note that
this force 
should, in principle, act on both proton and neutron (Feynman--Onsager) vortices.

The buoyancy force was derived in \cite{1985Ap&SS.115...43M} 
from the purely hydrodynamic consideration.
One may ask
whether it is legitimate
to apply such an approach
on microscopic length-scales comparable to the radius of the vortex core.
Another issue
is related to the fact that the buoyancy force
is always added in the equation of vortex motion as some `external' force.
This implies that it should not be contained, for example, 
in the smooth-averaged hydrodynamics 
of \cite{1961JETP..13..643B} (and its multifluid extensions).
But is it true? 

We address these issues in the present note.
The work is organized as follows.
Section \ref{buo} contains a brief discussion of the vortex structure, 
together with the
Muslimov \& Tsygan derivation of the buoyancy force, 
and alternative derivation leading to a different result.
The new expression for the buoyancy force is then compared 
to the results of microscopic calculations based on the solution
to the conservative Ginzburg--Landau (Gross--Pitaevsky) equation.
Section \ref{hvbk} clarifies whether the buoyancy effects 
are already present in the hydrodynamics of \cite{1961JETP..13..643B}.
Section \ref{appl} considers a few applications of our results.
Finally, we sum up in section~\ref{concl}.
For simplicity, we neglect the relativistic and entrainment effects 
in this note.
They are thoroughly discussed in \cite{2016arXiv160701629G}.

\section{Buoyancy force on a vortex}
\label{buo}

\subsection{Vortex structure}

Consider first a proton vortex. 
It consists of a non-superconducting core
with the radius of the order of 
the proton coherence length $\xi_{\rm p} \equiv \hbar \pFp/(\pi m_{\rm p}^\ast \Delta_{\rm p})$,
\begin{equation}
\xi_{\rm p}
	\approx 1.7 \times 10^{-12}\, {\rm cm} \left(\frac{n_{\rm p}}{0.18 n_0} \right)^{1/3}
	\left(\frac{m_{\rm p}}{m_{\rm p}^\ast}\right) \left(\frac{0.76 \, {\rm MeV}}{\Delta_{\rm p}}\right),
\label{eq:coh}
\end{equation}
surrounded by superconducting currents
with the velocity field (e.g., \citealt{degennes99,ll80})
\begin{gather}
\label{eq:v(r)}
	\v{v}(\v{r})
=  \frac{e_{\rm p} \phi_0}{2\pi m_{\rm p} c\, \delta_{\rm p}} K_1 \left( \frac{r}{\delta_{\rm p}} \right) \v{e}_\varphi
.
\end{gather}
Here 
$\pFp$, $n_{\rm p}$, $m_{\rm p}$, $m_{\rm p}^\ast$, $\Delta_{\rm p}$ 
are the proton Fermi momentum, number density, mass, effective mass, and energy gap, respectively;
$n_0=0.16$~fm$^{-3}$ is the nuclear matter density;
$r$ is the distance to the vortex core;
$\v{e}_\varphi$ is the unit vector in the azimuthal direction%
%
\footnote{
We use the cylindrical coordinate system $(r,~\varphi,~z)$
with the axis $z$ directed along the vortex line.};
%
and $K_i(x)$ is the MacDonald function.
Finally, 
$\delta_{\rm p}\equiv [m_{\rm p}^2 c^2/(4 \pi e_{\rm p}^2 \rho_{\rm sp})]^{1/2}$ 
is the London penetration depth for protons,
\begin{gather}
\delta_{\rm p} \approx 4.24 \times 10^{-12} \, {\rm cm} 
\left( \frac{0.18 n_0}{n_{\rm sp}} \right)^{1/2},
\label{dp}
\end{gather}
where $\rho_{\rm sp}$ and $n_{\rm sp}= \rho_{\rm sp}/m_{\rm p}$
are, respectively, the `superconducting' proton mass density and number density.
At $r \ll \delta_{\rm p}$ the velocity 
$v(r) \approx \kappa_{\rm p} / (2\pi r)$
[$\kappa_{\rm p} \equiv \pi \hbar / m_{\rm p}$ is the circulation quantum],
while at $r \gg \delta_{\rm p}$ it decays exponentially, 
$v(r) \propto {\rm exp}(-r/\delta_{\rm p})$.
The proton supercurrent produces a magnetic field,
\begin{gather}
	\v{B}(r)
	= \frac{\phi_0}{2\pi \delta_{\rm p}^2}
		K_0 \left(\frac{r}{\delta_{\rm p}}\right)
		{\pmb \nu}
,
\end{gather}
where ${\pmb \nu} \equiv \v{e}_z$ is the unit vector in the vortex direction 
(along $z$). 
The vortex energy per unit length, $\hat{E}_{\rm Vp}$,
is given by the formula (\citealt{ll80})
\begin{gather}
\label{eq:Evp}
	\hat{E}_{\rm Vp}
		=
			\int
				\left[
					\rho_{\rm sp} \frac{v^2(r)}{2}
					+ \frac{B^2(r)}{8\pi}
				\right]
				r {\rm d} r {\rm d} \varphi
		\approx		
			\rho_{\rm sp} \frac{\kappa_{\rm p}^2}{4\pi} \ln \frac{\delta_{\rm p}}{\xi_{\rm p}}
.
\end{gather}
To allow for the entrainment effect, 
$\rho_{\rm sp}$ in equation \eqref{eq:Evp} 
should be replaced with 
the element $\rho_{\rm pp}$ of the entrainment matrix $\rho_{ik}$ \citep{mendell91a}.

Similar formulas can also be written for vortices in an uncharged superfluids, 
e.g., for neutron vortex%
%
\footnote{In this paper we assume, for simplicity, 
that neutrons pair in the spin-singlet ($^1$S$_0$) state.}.
%
Its velocity field is
$\v{v}(\v{r}) = \kappa_{\rm n}/({2\pi r}) \v{e}_\varphi$,
and energy per unit length
\begin{gather}
\label{eq:Evn}
\hat{E}_{\rm Vn}
\approx		
\rho_{\rm sn} \frac{\kappa_{\rm n}^2}{4\pi} \ln \frac{b_{\rm n}}{\xi_{\rm n}}
,
\end{gather}
where $\kappa_{\rm n} \equiv \pi\hbar / m_{\rm n}$;
$\xi_{\rm n}$ is the neutron coherence length, given by the formula 
similar to equation \eqref{eq:coh};
and $b_{\rm n}$ is some `external' radius of the order of the typical
length-scale of the problem
(e.g., intervortex spacing or radius of a vessel).
To account for the entrainment, 
$\rho_{\rm sn}$ in equation \eqref{eq:Evn} 
should be replaced 
with the combination $(\rho_{\rm nn}\rho_{\rm pp}-\rho_{\rm np}^2)/\rho_{\rm pp}$
(\citealt{mendell91a,2011MNRAS.410..805G}).

Note that in inhomogeneous medium 
(when $\rho_{\rm sn}$ slowly varies in space)
the neutron vortex velocity
differs from
$\v{v}(\v{r}) = \kappa_{\rm n}/({2\pi r}) \v{e}_\varphi$ 
\citep[see, e.g.,][]{1994PhyD...78....1R,sr04},
but the vortex energy $\hat{E}_{\rm Vn}$
still has the same form~\eqref{eq:Evn}
up to the terms 
quadratic
in $|\gv{\nabla} \rho_{\rm sn}|$.
The same, of course, applies to proton vortices.

\subsection{Buoyancy force: standard derivation}
\label{stand}

The concept of the buoyancy force on a vortex was introduced 
by \citet{1985Ap&SS.115...43M} \citep[see also][]{hrs86}
who considered proton fluxtubes in superconducting interiors of neutron stars.
These authors noted that in the vicinity of the vortex core
the fluid pressure (and, therefore, density) decreases
because of 
local magnetic field and supercurrents associated with the vortex.
The pressure drop $\Delta P(r)$ at a distance $r$ from the vortex line 
is given by
\begin{gather}
	\Delta P(r)
		=  \rho_{\rm sp} \frac{v^2(r)}{2} + \frac{B^2(r)}{8\pi}
.
\label{Pres}
\end{gather}
Therefore, the buoyancy force per unit length $\v{f}_B$,
arising due to the density drop $\Delta\rho (r)$,
equals
\begin{gather}
\label{eq:fB-1985}
	\v{f}_B
		= - \v{g} \int_{\xi_{\rm p}}^{\infty} \Delta\rho(r) r {\rm d}r {\rm d} \varphi
		= - \hat{E}_{\rm Vp} \frac{\v{g}}{c_{\rm s}^2}
,
\end{gather}
where $\v{g}$ is the local gravitational acceleration.
Here we used the relation $\Delta \rho = c_s^{-2} \Delta P$,
where $c_{\rm s}^2 \equiv {\rm d}P / {\rm d} \rho$ is the squared speed of sound.
Using the hydrostatic equilibrium condition
$\rho \v{g}  = \gv{\nabla} P$,
one can rewrite the buoyancy force as
\begin{gather}
\label{eq:f-grad-rho}
	\v{f}_B
		= - \hat{E}_{\rm Vp} \frac{\v{g}}{c_{\rm s}^2}
		= - \hat{E}_{\rm Vp} \frac{\gv{\nabla}\rho}{\rho}
.
\end{gather}
The proposed simple 
derivation predicts that there is a non-zero 
force acting on a vortex in an external gravitational field.
One may see that this force is derived from the purely `hydrodynamic' arguments.
However, the validity of such a hydrodynamic approach 
is questionable
at microscopic length-scales 
(e.g., use of the notion of pressure 
in the very vicinity of the vortex core looks doubtful).
Below we demonstrate that
equation \eqref{eq:f-grad-rho} is correct only in one special case 
and, generally, it should be replaced by a different formula.

\subsection{Alternative derivation and different result}
\label{sec:fB-2}

The force acting on a vortex in inhomogeneous medium
can also be derived from simple energetic arguments.
Assume first that the vortex is directed along the axis $z$,
while 
the background superfluid density $\rho_{\rm sp}$ depends on $x$ only 
[here and below we use the standard Cartesian coordinates $(x,\, y,\, z)$].
Now, if we shift the vortex from $x$ to $x'=x+\delta x$,
its energy \eqref{eq:Evp} will change according to
\begin{gather}
	\hat{E}_{\rm Vp} (x') - \hat{E}_{\rm Vp} (x)
		\approx \left[ \rho_{\rm sp}(x') - \rho_{\rm sp}(x) \right]
			\frac{\kappa_{\rm p}^2}{4\pi} \ln \frac{\delta_{\rm p}}{\xi_{\rm p}}
,
\end{gather}
so that the buoyancy force,
given by the (minus) vortex energy gradient, will be 
$\v{f}_B = - \gv{\nabla} \hat{E}_{\rm Vp} (\v{r})$.
In the more general case, when $\gv{\nabla} \rho_{\rm sp}$ 
is not necessarily directed along $x$, one has
\begin{gather}
\label{eq:f-grad-rho_s}
	\v{f}_B
		= - \gv{\nabla}_\perp \hat{E}_{\rm Vp} (\v{r})
		= - \hat{E}_{\rm Vp}  \frac{\gv{\nabla}_\perp \rho_{\rm sp}}{\rho_{\rm sp}}
,
\end{gather}
where $\gv{\nabla}_\perp \equiv \gv{\nabla} - \v{\nu} (\v{\nu} \gv{\nabla})$
is the component of gradient perpendicular
to the vortex line direction $\v{\nu}$.
The appearance of $\gv{\nabla}_\perp$ instead of $\gv{\nabla}$ in this formula
reflects the fact that the 
translation of the vortex along the axis $z$ does not change its energy.
Of course, similar formula (with index $p$ replaced by $n$) 
also applies to neutron (Feynman--Onsager) vortices 
(\citealt{sr04}; see also section \ref{rubi}).

Generally, the formula \eqref{eq:f-grad-rho_s} 
differs from the `classical' result \eqref{eq:f-grad-rho}.
They are equivalent only for a one-component liquid in the hydrostatic equilibrium 
and at zero temperature ($T=0$), when $\rho_{\rm sp} = \rho$.

\subsection{Comparison with the microscopic theory}
\label{rubi}

It is interesting to confront our formula \eqref{eq:f-grad-rho_s}
with the results of microscopic theory.
For example, \cite{1994PhyD...78....1R}
analysed, within the conservative Ginzburg--Landau (Gross--Pitaevsky) equation, 
the motion of a neutral (e.g., neutron) vortex in the system 
at $T=0$ 
with a slightly inhomogeneous background superfluid density $\rho_{\rm sn}(x,y)$. 
They considered a situation in which
the vortex is straight and directed along the axis $z$.
In such formulation the problem is two-dimensional -- 
the vortex moves in the $xy$-plane.
Using the method of matched asymptotic expansion, 
\cite{1994PhyD...78....1R}
found the following analytical expression for the vortex velocity $\v{V}_{\rm Ln}$:
\begin{gather}
\label{eq:VL-micro}
\v{V}_{\rm Ln}
= \v{V}_{\rm sn}
- \frac{\kappa_{\rm n}}{4\pi} \,
\v{\nu}
\times
\frac{\gv{\nabla} \rho_{\rm sn}}{\rho_{\rm sn}}
\ln \frac{R_{\rm n}}{\xi_{\rm n}}
.
\end{gather}
Here $\v{V}_{\rm sn}$ is the background superfluid velocity
and $R_{\rm n}$ is a typical length-scale of superfluid density variation,
\begin{gather}
R_{\rm n} = 2a_1 e^{1/2 - C}
\bigg/
{	
	\sqrt{
		\left(
		\frac{\gv{\nabla} \rho_{\rm sn}}{2 \rho_{\rm sn}}
		\right)^2
		- \frac{1}{2} \nabla^2 \ln \rho_{\rm sn}
	}
}
\end{gather}
where $C \approx 0.577$ is the Euler constant
and $\ln a_1 \approx 0.405$.

One can try to derive an equation similar to \eqref{eq:VL-micro} 
from the formula \eqref{eq:f-grad-rho_s} (see also \citealt{sr04}).
The only forces acting on a vortex in the conservative 
2D problem considered by \cite{1994PhyD...78....1R} are the Magnus force, 
$\v{f}_{\rm M}= \rho_{\rm sn} \kappa_{\rm n}\v{\nu}\times(\v{V}_{\rm Ln}-\v{V}_{\rm sn})$,
and the buoyancy force, $\v{f}_{\rm B}$.
Neglecting vortex inertia, one should have $\v{f}_{\rm B}+\v{f}_{\rm M}=0$,
which gives
%
\footnote{Note that $\v{V}_{\rm Ln}$ is defined
up to an arbitrary term parallel to the vortex line.
Such term 
does not affect the vortex dynamics (see, e.g., \citealt{khalatnikov00})
and is chosen here such that $\v{V}_{\rm Ln}\v{\nu}=\v{V}_{\rm sn}\v{\nu}$.
},
%
\begin{gather}
\label{eq:VLp}
\v{V}_{\rm Ln}
= \v{V}_{\rm sn}
- \frac{\hat{E}_{\rm Vn}}{\rho_{\rm sn}\kappa_{\rm n}}
\v{\nu} \times \frac{\gv{\nabla} \rho_{\rm sn}}{\rho_{\rm sn}}=
\v{V}_{\rm sn}- \frac{\kappa_{\rm n}}{4\pi} \,
\v{\nu}
\times
\frac{\gv{\nabla} \rho_{\rm sn}}{\rho_{\rm sn}}
\ln \frac{b_{\rm n}}{\xi_{\rm n}}
.
\end{gather}
This equation has exactly the same structure as \eqref{eq:VL-micro}
provided that we identify $b_{\rm n}=R_{\rm n}$, which is natural,
since this is the only `external' length-scale in the problem.
It follows from \eqref{eq:VLp} that in the absence of background superfluid velocity 
($\v{V}_{\rm sn}=0$)
vortex moves along the surface $\rho_{\rm sn}={\rm const}$, 
i.e. its energy is conserved. 

\section{Is there a buoyancy force in superfluid hydrodynamics?}
\label{hvbk}

In the context of neutron stars
the vortex buoyancy 
(in the form \ref{eq:fB-1985}) 
was routinely introduced 
as some external force (but see \citealt{jones06})
in order to explain 
expulsion of the magnetic field (confined to Abrikosov vortices) 
from the superconducting interiors of neutron stars.
Here we shall demonstrate that this force is 
absent in the often used version (\citealt{1960AdPhy...9...89H}) 
of superfluid hydrodynamics of \cite{1956RSPSA.238..215H} 
(see also a monograph by \citealt{2005qvhi.book.....D})
but is contained in the hydrodynamics
of \cite{1961JETP..13..643B} (and its various extensions).
Thus, introduction of this force `by hands' may lead to double-counting.
Note that, the fact that the buoyancy 
is already incorporated
in the hydrodynamic equations of \cite{1961JETP..13..643B},
has not been well recognized in the literature
(see, e.g., p.~94 of the authoritative review by \citealt{sonin87}, 
where this force is described as `external').

Consider a neutral superfluid (e.g., neutron liquid) 
containing 
vortices moving 
with some local velocity $\v{V}_{\rm Ln}$.
For simplicity, we shall be interested in the so called 
weak-drag limit, when the mutual friction forces can be neglected 
(note that, effectively, \citealt{1994PhyD...78....1R} 
also worked in that limit, see section \ref{rubi}).
In this approximation the hydrodynamics of \cite{1961JETP..13..643B} 
gives the following expression for $\v{V}_{\rm Ln}$ 
(see equation 16.45 in the monograph by \citealt{khalatnikov00}
with $\beta=\beta^\prime=0$
%
\footnote{Note a misprint in this equation: 
the second term in its r.h.s.\ 
should have `+' sign in front of $1/\rho_{\rm s} \, {\rm curl} \,  \lambda \v{\nu}$.}),
%
\begin{gather}
\v{V}_{\rm Ln}^{\rm (BK)}
= \v{V}_{\rm sn} + \frac{1}{\rho_{\rm sn}} {\rm curl} \lambda_{\rm n} \v{\nu}
=\v{V}_{\rm sn} + \frac{1}{\rho_{\rm sn}} 
{\rm curl}\left(\frac{\hat{E}_{\rm Vn}}{\kappa_{\rm n}} \v{\nu}\right)
\nonumber\\
= \v{V}_{\rm sn} 
-\frac{\hat{E}_{\rm Vn}}{\kappa_{\rm n}\rho_{\rm sn}}\, 
\v{\nu}\times \frac{\v{\nabla} \rho_{\rm sn}}{\rho_{\rm sn}}
+\frac{\hat{E}_{\rm Vn}}{\kappa_{\rm n}\rho_{\rm sn}}\, {\rm curl}\, \v{\nu}
,
\label{eq:VL-BK-1}
\end{gather}
where $\lambda_{\rm n} \equiv \hat{E}_{\rm Vn} / \kappa_{\rm n}$ 
and in the second line we make use of the fact that $\hat{E}_{\rm Vn}$ is 
approximately proportional to $\rho_{\rm sn}$ [see equation \eqref{eq:Evn}].
In turn, the hydrodynamics of \cite{1960AdPhy...9...89H}
predicts that in the weak-drag limit $\v{V}_{\rm Ln}$ equals
\begin{gather}
\label{eq:VL-Hall}
\v{V}_{\rm Ln}^{\rm (H)}
= \v{V}_{\rm sn} +\frac{\hat{E}_{\rm Vn}}{\kappa_{\rm n}\rho_{\rm sn}}\, {\rm curl}\, \v{\nu}
.
\end{gather}
Let us compare equations \eqref{eq:VL-BK-1}, \eqref{eq:VL-Hall},
and equation \eqref{eq:VLp} from the previous section.
First of all, 
the last terms in the r.h.s.\ of equations \eqref{eq:VL-BK-1} and \eqref{eq:VL-Hall}
coincide and appear due to the local vortex curvature (and related tension), which
generates small contribution to $\v{V}_{\rm Ln}$,
making it a bit different from the average `transport' velocity $\v{V}_{\rm sn}$
(see \citealt{2005qvhi.book.....D} for details;
note, however, that the corresponding discussion in that reference
contains a number of misprints).
Because this term vanishes for straight vortices, 
it does not appear in equation \eqref{eq:VLp}.

Next, the second term in the r.h.s. of equation
\eqref{eq:VL-BK-1} indicates that 
the difference between the vortex velocity $\v{V}_{\rm Ln}$
and the transport velocity $\v{V}_{\rm sn}$
is driven
not only by vortex curvature, 
but also by inhomogeneity of the background, $\gv{\nabla} \rho_{\rm sn}$.
It exactly coincides with the similar `buoyancy' term in equation \eqref{eq:VLp},
but is absent in the Hall's hydrodynamics [see equation \eqref{eq:VL-Hall}].

We come to conclusion that the superfluid hydrodynamics
of \cite{1961JETP..13..643B} implicitly contains 
all 
buoyancy effects,
whereas one has to include the buoyancy force `by hands'
if one prefers to use the superfluid hydrodynamics of \cite{1960AdPhy...9...89H}.

It remains to note
that the buoyancy effects
are automatically included 
in both non-relativistic and relativistic 
extensions of Bekarevich and Khalatnikov hydrodynamics,  
describing 
superfluid/superconducting mixtures
(see, e.g., \citealt{1991AnPhy.205..110M,mendell91a,ss95,2011MNRAS.410..805G,
2016PhRvD..93f4033G,2016arXiv160701629G}). 

\section{Some applications}
\label{appl}

\subsection{Vortices in a rotating neutral superfluid}

Let us
discuss, within the hydrodynamics of \cite{1961JETP..13..643B}, 
the equilibrium configuration of a rotating neutral superfluid with non-zero $\v{\nabla}\rho_{\rm sn}$.
For simplicity, we restrict ourselves to the cylindrical geometry 
(e.g., cylindrical vessel rotating at a frequency $\v{\Omega}$
around its symmetry axis), 
and consider the case 
of a one-component liquid at $T=0$ 
(hence $\rho_{\rm sn}$ equals 
the total mass density $\rho_{\rm n}$).
Generalization of these results to finite temperatures is straightforward
and does not change our main conclusions.

The superfluid is governed by the `superfluid' and continuity equations 
(see equations 16.24 and 16.40 in \citealt{1961JETP..13..643B}), 
which can be rewritten in the frame, rotating with the vessel, as
\begin{gather}
\frac{\partial \tilde{\v{V}}_{\rm sn}}{\partial t}+
2 \, [\v{\Omega}\times \tilde{\v{V}}_{\rm sn}]
+(\tilde{\v{V}}_{\rm sn}\v{\nabla}) \tilde{\v{V}}_{\rm sn}
+\v{\nabla}(\breve{\mu}+\phi)
\nonumber\\
=-\frac{1}{\rho_{\rm sn}}[\rm{curl} \, \tilde{\v{V}}_{\rm sn} 
\times {\rm curl}(\lambda_{\rm n}\v{\nu})]
-\frac{1}{\rho_{\rm sn}}[2 \v{\Omega} 
\times {\rm curl}(\lambda_{\rm n}\v{\nu})]
\label{sfl1}\\
\frac{\partial \rho_{\rm sn}}{\partial t}+{\rm div} (\rho_{\rm sn} \tilde{\v{V}}_{\rm sn})=0,
\label{sfl2}
\end{gather}
where 
$\tilde{\v{V}}_{\rm sn}=\v{V}_{\rm sn}-[\v{\Omega}\times \v{r}]$ is measured in the rotating frame;
$\breve{\mu}$ is the chemical potential per unit mass;
and
$\phi$ is the sum of centrifugal potential
plus potential of some external force producing the density gradient
(in neutron stars it is the gravitational potential).
In the absence of density gradients, 
when the right-hand side
of equation \eqref{sfl1} vanishes%
%
\footnote{Note that $\rm{curl}\, \v{\nu}=0$ because $\v{\nu}$ is collinear with $\v{\Omega}$.},
%
one has $\tilde{\v{V}}_{\rm sn}=0$ in equilibrium, 
that is superfluid mimics solid-body rotation.
However, when the density gradients are allowed for, 
the solution to equations \eqref{sfl1} and \eqref{sfl2} describes differential rotation,
\begin{gather}
\tilde{\v{V}}_{\rm sn}=\frac{\lambda_{\rm n}}{\rho_{\rm sn}} \, \v{\nu}
\times\frac{\v{\nabla}\rho_{\rm sn}}{\rho_{\rm sn}}.
\label{sol1}
\end{gather}
At the same time, using \eqref{eq:VL-BK-1}
one obtains that 
vortices move with the vessel 
(and with the normal component if $T\neq 0$):
$\v{V}_{\rm Ln}=[\v{\Omega}\times \v{r}]$.
For neutron stars the typical precession 
frequency of vortices with respect to the superfluid component, 
$\omega_{\rm P} \equiv |\tilde{\v{V}}_{\rm sn}|/r 
=\kappa_{\rm n}/(4\pi r) \, \ln(b_{\rm n}/\xi_{\rm n}) 
\, |\gv{\nabla}({\rm ln}\rho_{\rm sn})|$,
is extremely small, $\omega_{\rm P}\sim 10^{-15}$~s$^{-1}$,
but it is much larger (and is observed) 
in Bose--Einstein condensates (\citealt{ahwc00}).
It is interesting that a similar formula for $\omega_{\rm P}$
has been obtained in \cite{sr04} by different means.

\subsection{Magnetic field evolution in superconducting neutron-star cores}

If protons in a neutron star core form a type-II superconductor,
then almost all magnetic field
is confined to proton vortices (e.g., \citealt{2011MNRAS.410..805G}).
Therefore, evolution of the 
magnetic induction $\v{B}$ in the superconducting region
is governed by the equation \citep[e.g.,][]{kg01a}, 
\begin{gather}
\label{eq:dBdt}
	\pd{\v{B}}{t}
	= \rot \left[ \v{V}_{\rm Lp} \times \v{B} \right],
\end{gather}
which describes transport of $\v{B}$ with the velocity $\v{V}_{\rm Lp}$ of proton vortices.
The latter is given by the expression
\begin{equation}
{\pmb V}_{{\rm Lp}} = 
{\pmb V}_{\rm norm} - \alpha_{\rm p} \, 
{\pmb W}_{\rm p} - \beta_{\rm p} \,
{\pmb \nu}\times {\pmb W}_{\rm p}, 
\label{Vlapp}
\end{equation}
where $\v{W}_{\rm p}$ can be presented as a sum of two terms related, respectively,
to buoyancy and vortex tension,
\begin{gather}
{\pmb W}_{\rm p} =
{\rm curl} \left( \lambda_{\rm p}\v{\nu}\right)
= -\frac{\hat{E}_{\rm Vp}}{\kappa_{\rm p}}\, 
\v{\nu}\times \frac{\v{\nabla} \rho_{\rm sp}}{\rho_{\rm sp}}
+\frac{\hat{E}_{\rm Vp}}{\kappa_{\rm p}}\, {\rm curl}\, \v{\nu}.
\label{Wp}
\end{gather}
In equations \eqref{Vlapp} and \eqref{Wp}
$\lambda_{\rm p}=\hat{E}_{\rm Vp}/\kappa_{\rm p}$;
$\v{\nu}=\v{B}/B$;
$\alpha_{\rm p}$ and $\beta_{\rm p}$
are the (poorly known) mutual friction coefficients.
These equations have been recently derived in \cite{2016arXiv160701629G}
under assumption that all the normal (non-superfluid and non-superconducting) liquid components
move with one and the same `normal' velocity $\v{V}_{\rm norm}$ 
%
\footnote{
\label{foot}	
\cite{2015MNRAS.453..671G} proposed a magnetic field evolution equation
which is different from \eqref{Vlapp}; we discuss it below.
}.
%
This assumption can be relaxed (Gusakov \& Dommes, in preparation),
which modifies $\v{V}_{\rm Lp}$ by adding 
a diffusion-induced term to $\v{W}_{\rm p}$. 

It is instructive to estimate
typical time-scales appearing in the evolution equation \eqref{eq:dBdt}.
Following \cite{2015MNRAS.453..671G}, one gets
\begin{gather}
	\tau_{\rm cons}
	= \frac{\kappa_{\rm p}L^2}{|\alpha_{\rm p}|\hat{E}_{\rm Vp}}
	=\frac{4 \pi L^2}{|\alpha_{\rm p}| \rho_{\rm sp}\kappa_{\rm p}}\,
		\left[
			{\rm ln } \left(\frac{\delta_{\rm p}}{\xi_{\rm p}}\right)
		\right]^{-1},
\label{taucons}
\\
\tau_{\rm diss}
	= \frac{\kappa_{\rm p}L^2}{\beta_{\rm p}\hat{E}_{\rm Vp}}
	=\frac{4 \pi L^2}{\beta_{\rm p} \rho_{\rm sp}\kappa_{\rm p}}\,
		\left[{\rm ln }
			\left(\frac{\delta_{\rm p}}{\xi_{\rm p}}\right)
		\right]^{-1}
.
\label{taudiss}
\end{gather}
Here $L$ is the characteristic length-scale of the problem,
characterizing variation of
the magnetic induction $\v{B}$, or superfluid density $\rho_{\rm sp}$, 
or vortex direction $\v{\nu}$.
Note that, near the normal-superconducting boundary, 
the ratio $\v{\nabla} \rho_{\rm sp}/\rho_{\rm sp}$ 
[see the `buoyancy' term in equation \eqref{Wp}] can be very large, 
since there $\rho_{\rm sp}\rightarrow 0$, 
while $\v{\nabla} \rho_{\rm sp}$ is finite.
Also, $\v{\nabla} \rho_{\rm sp}$ can be large (i.e., $L$ is small) 
near the crust-core interface
and, generally, in the vicinity of any place in the core
where the superfluid proton density 
varies sharply.
%
\footnote{We thank A.I.~Chugunov for pointing out to us this possibility.}

The time-scales \eqref{taucons} and \eqref{taudiss} are called, respectively, 
`conservative' and `dissipative'
because it can be shown (e.g., \citealt{1991AnPhy.205..110M,2016arXiv160701629G}) 
that the terms $\propto \beta_{\rm p}$ create entropy in the system, while 
those $\propto \alpha_{\rm p}$ do not.
The mutual friction coefficients $\alpha_{\rm p}$ and $\beta_{\rm p}$
are rather uncertain, especially at temperatures comparable 
to the nucleon critical temperatures,
and there are no agreement in the literature on their actual values 
(e.g., \citealt{als84, jones06, asc06, jones09}).
At $T=0$ these coefficients
can be expressed through the drag coefficient $\mathcal{R}_{\rm p} \sim 2 \times 10^{-4}$
\citep{2015MNRAS.453..671G} as
\begin{gather}
\label{eq:alpha-beta-R}
	\alpha_{\rm p}
	= - \frac{1}{\rho_{\rm sp}} \frac{1}{1 + \mathcal{R}_{\rm p}^2}
,\quad
	\beta_{\rm p}
	= \frac{1}{\rho_{\rm sp}} \frac{\mathcal{R}_{\rm p}}{1 + \mathcal{R}_{\rm p}^2}
.
\end{gather}
Adopting these values, one obtains, to leading order in $\mathcal{R}_{\rm p}$,
\begin{gather}
	\tau_{\rm cons} \approx 2 \times 10^8 L_6^2 \,\, {\rm yr},
\label{taucons1}
\quad
	\tau_{\rm diss} \approx 2 \times 10^{12} L_6^2 
	\left(\frac{\mathcal{R}_{\rm p}}{10^{-4}}\right)^{-1} \,\, {\rm yr}
,
\end{gather}
where $L_6 = L / (10^6~{\rm cm})$.
Clearly, these estimates should be taken with caution.
Still, they indicate that the time-scale $\tau_{\rm cons}$
can be comparable to the typical time-scale of magnetic field evolution in pulsars.

Note that $\tau_{\rm cons}$ 
is smaller than
the corresponding estimate in
\cite{2015MNRAS.453..671G} by 7 orders of magnitude, 
which is a consequence of a bit different evolution equation 
used by these authors.
The evolution equation 
of
\cite{2015MNRAS.453..671G} (their equation 67) reads
\footnote{It also immediately follows from
equations (160) and (162) of \cite{2011MNRAS.410..805G}.}
\begin{gather}
\label{eq:dBdt-Graber}
	\pd{\v{B}}{t}
	= \rot \left[ \left( 
		\v{V}_{\rm Lp} - \underline{\frac{\hat{E}_{\rm Vp}}{\rho_{\rm sp} \kappa_{\rm p}}\, {\rm curl}\, \v{\nu}}
		\right)\times \v{B} \right]
,
\end{gather}
where the expression for the vortex velocity,
\begin{gather}
\label{eq:VLp-Graber}
	\v{V}_{\rm Lp}
	= \v{V}_{\rm norm}
		+ \frac{\hat{E}_{\rm Vp}}{\rho_{\rm sp} \kappa_{\rm p} (1 + \mathcal{R}_{\rm p}^2)}
			\left(
			\rot \v{\nu}
			- \mathcal{R}_{\rm p} \v{\nu} \times \rot \v{\nu}
			\right)
,
\end{gather}
coincides with equation~\eqref{Vlapp}
if 
$\alpha_{\rm p}$ and $\beta_{\rm p}$ are given by equation~\eqref{eq:alpha-beta-R}
and, additionally, $\v{\nabla} \rho_{\rm sp} = 0$
(which means that
the authors
ignore the buoyancy force).
Equation \eqref{eq:dBdt-Graber} is rather puzzling:
it states that the magnetic field is transported with the velocity
that \emph{differs} from the vortex velocity $\v{V}_{\rm Lp}$,
in contradiction to the (explicitly made) assumption 
that all the magnetic field is stored in vortices.
Note that this equation
has been used subsequently by \cite{2016MNRAS.456.4461E} 
for detailed modelling of the magnetic field evolution
in neutron stars (see equation 16 in that reference).
These authors also accounted for the buoyancy force,
but in the incorrect form~\eqref{eq:f-grad-rho}.

Now the reason for the seven-order difference 
in the estimates for $\tau_{\rm cons}$
can be easily pinpointed.
To this aim we expand the second term in equation (\ref{eq:VLp-Graber}) 
in the Taylor series in small parameter $\mathcal{R}_{\rm p}$.
The leading-order contribution in that expansion
cancels out the underlined term in equation (\ref{eq:dBdt-Graber}).
As a result, $\tau_{\rm cons}$
in \cite{2015MNRAS.453..671G}
is larger than our estimate
\eqref{taucons}
by a factor $\sim \mathcal{R}_{\rm p}^{-2} \sim 10^7$.

\section{Summary}
\label{concl}

Our results can be summarized as follows:

\begin{enumerate}

\item 
The standard expression \eqref{eq:f-grad-rho} for the buoyancy force
acting on a vortex is generally incorrect
and should be replaced with equation~\eqref{eq:f-grad-rho_s}.
Essentially, one has to replace in equation 
\eqref{eq:f-grad-rho}
$\v{g}/c_{\rm s}^2  = \gv{\nabla} \rho / \rho$
with
$\gv{\nabla}_\perp \rho_{\rm sp} / \rho_{\rm sp}$
(for proton vortices),
or with $\gv{\nabla}_\perp \rho_{\rm sn} / \rho_{\rm sn}$ (for neutron vortices).
In some regions of a neutron star this can
significantly increase the buoyancy force
(and may even reverse its direction).
In particular, buoyancy effects can be important
near the boundary of superconducting region,
where $\rho_{\rm sp}$ varies sharply.

\item
The proposed buoyancy force derivation
is supported by microscopic calculations 
within the conservative Ginzburg--Landau 
(Gross--Pitaevsky) equation
(\citealt{1994PhyD...78....1R}; see also \citealt{sr04}).

\item
The same buoyancy force is already implicitly
accounted for in the superfluid hydrodynamics
of \citet{1961JETP..13..643B}
and its extensions to
superfluid/superconducting mixtures
\citep{1991AnPhy.205..110M, mendell91a, ss95, 2011MNRAS.410..805G, 2016arXiv160701629G}.
To the best of our knowledge,
the corresponding terms 
in the equations describing superfluids
have not been interpreted as `buoyancy-related'
(see, e.g., p.~94 of the review by \citealt{sonin87}, 
where 
the buoyancy force
is explicitly considered as `external').

\item
However, this force 
does not appear in superfluid hydrodynamics of \citet{1960AdPhy...9...89H} 
\cite[see also][]{2005qvhi.book.....D},
which is equivalent to that of Bekarevich \& Khalatnikov in (almost) all other aspects.

\item We briefly discussed a number of applications in
which the buoyancy force may lead to noticeable effects.
These include hydrostatic equilibrium 
of neutral rotating superfluids,
as well as the magnetic field evolution in superconducting neutron stars.
Our estimates indicate that the magnetic field can
evolve 
on a much shorter timescale
than it was assumed by \cite{2015MNRAS.453..671G}
(and, subsequently, by \citealt{2016MNRAS.456.4461E}),
who used a different (and incorrect, as we argue in Section 4) magnetic field evolution equation.

\end{enumerate}

\section{Acknowledgements}

We are very grateful to Elena Kantor, Andrey Chugunov, Kostas Glampedakis,
and D.A. Shalybkov
for valuable discussions and comments.
This study was supported by the Russian Science Foundation (grant number 14-12-00316).

\bibliography{litt}

\begin{thebibliography}{}
\makeatletter
\relax
\def\mn@urlcharsother{\let\do\@makeother \do\$\do\&\do\#\do\^\do\_\do\%\do\~}
\def\mn@doi{\begingroup\mn@urlcharsother \@ifnextchar [ {\mn@doi@}
  {\mn@doi@[]}}
\def\mn@doi@[#1]#2{\def\@tempa{#1}\ifx\@tempa\@empty \href
  {http://dx.doi.org/#2} {doi:#2}\else \href {http://dx.doi.org/#2} {#1}\fi
  \endgroup}
\def\mn@eprint#1#2{\mn@eprint@#1:#2::\@nil}
\def\mn@eprint@arXiv#1{\href {http://arxiv.org/abs/#1} {{\tt arXiv:#1}}}
\def\mn@eprint@dblp#1{\href {http://dblp.uni-trier.de/rec/bibtex/#1.xml}
  {dblp:#1}}
\def\mn@eprint@#1:#2:#3:#4\@nil{\def\@tempa {#1}\def\@tempb {#2}\def\@tempc
  {#3}\ifx \@tempc \@empty \let \@tempc \@tempb \let \@tempb \@tempa \fi \ifx
  \@tempb \@empty \def\@tempb {arXiv}\fi \@ifundefined
  {mn@eprint@\@tempb}{\@tempb:\@tempc}{\expandafter \expandafter \csname
  mn@eprint@\@tempb\endcsname \expandafter{\@tempc}}}

\bibitem[\protect\citeauthoryear{{Alpar}, {Langer}  \& {Sauls}}{{Alpar}
  et~al.}{1984}]{als84}
{Alpar} M.~A.,  {Langer} S.~A.,   {Sauls} J.~A.,  1984, \mn@doi [\apj]
  {10.1086/162232}, \href {http://adsabs.harvard.edu/abs/1984ApJ...282..533A}
  {282, 533}

\bibitem[\protect\citeauthoryear{{Anderson}, {Haljan}, {Wieman}  \&
  {Cornell}}{{Anderson} et~al.}{2000}]{ahwc00}
{Anderson} B.~P.,  {Haljan} P.~C.,  {Wieman} C.~E.,   {Cornell} E.~A.,  2000,
  \mn@doi [Physical Review Letters] {10.1103/PhysRevLett.85.2857}, \href
  {http://adsabs.harvard.edu/abs/2000PhRvL..85.2857A} {85, 2857}

\bibitem[\protect\citeauthoryear{{Andersson}, {Sidery}  \& {Comer}}{{Andersson}
  et~al.}{2006}]{asc06}
{Andersson} N.,  {Sidery} T.,   {Comer} G.~L.,  2006, \mn@doi [\mnras]
  {10.1111/j.1365-2966.2006.10147.x}, \href
  {http://adsabs.harvard.edu/abs/2006MNRAS.368..162A} {368, 162}

\bibitem[\protect\citeauthoryear{{Bekarevich} \& {Khalatnikov}}{{Bekarevich} \&
  {Khalatnikov}}{1961}]{1961JETP..13..643B}
{Bekarevich} I.~L.,  {Khalatnikov} I.~M.,  1961, JETP, 13, 643

\bibitem[\protect\citeauthoryear{{Chau}}{{Chau}}{1997}]{chau97}
{Chau} H.~F.,  1997, \apj, \href
  {http://adsabs.harvard.edu/abs/1997ApJ...479..886C} {479, 886}

\bibitem[\protect\citeauthoryear{{Chau}, {Cheng}  \& {Ding}}{{Chau}
  et~al.}{1992}]{ccd92}
{Chau} H.~F.,  {Cheng} K.~S.,   {Ding} K.~Y.,  1992, \mn@doi [\apj]
  {10.1086/171917}, \href {http://adsabs.harvard.edu/abs/1992ApJ...399..213C}
  {399, 213}

\bibitem[\protect\citeauthoryear{De~Gennes}{De~Gennes}{1999}]{degennes99}
De~Gennes P.,  1999, Superconductivity Of Metals And Alloys.
Advanced Books Classics Series, Westview Press

\bibitem[\protect\citeauthoryear{{Ding}, {Cheng}  \& {Chau}}{{Ding}
  et~al.}{1993}]{dcc93}
{Ding} K.~Y.,  {Cheng} K.~S.,   {Chau} H.~F.,  1993, \mn@doi [\apj]
  {10.1086/172577}, \href {http://adsabs.harvard.edu/abs/1993ApJ...408..167D}
  {408, 167}

\bibitem[\protect\citeauthoryear{{Donnelly}}{{Donnelly}}{2005}]{2005qvhi.book.....D}
{Donnelly} R.~J.,  2005, {Quantized Vortices in Helium II}

\bibitem[\protect\citeauthoryear{{Elfritz}, {Pons}, {Rea}, {Glampedakis}  \&
  {Vigan{\`o}}}{{Elfritz} et~al.}{2016}]{2016MNRAS.456.4461E}
{Elfritz} J.~G.,  {Pons} J.~A.,  {Rea} N.,  {Glampedakis} K.,   {Vigan{\`o}}
  D.,  2016, \mn@doi [\mnras] {10.1093/mnras/stv2963}, \href
  {http://adsabs.harvard.edu/abs/2016MNRAS.456.4461E} {456, 4461}

\bibitem[\protect\citeauthoryear{{Glampedakis}, {Andersson}  \&
  {Samuelsson}}{{Glampedakis} et~al.}{2011}]{2011MNRAS.410..805G}
{Glampedakis} K.,  {Andersson} N.,   {Samuelsson} L.,  2011, \mn@doi [\mnras]
  {10.1111/j.1365-2966.2010.17484.x}, \href
  {http://adsabs.harvard.edu/abs/2011MNRAS.410..805G} {410, 805}

\bibitem[\protect\citeauthoryear{{Graber}, {Andersson}, {Glampedakis}  \&
  {Lander}}{{Graber} et~al.}{2015}]{2015MNRAS.453..671G}
{Graber} V.,  {Andersson} N.,  {Glampedakis} K.,   {Lander} S.~K.,  2015,
  \mn@doi [\mnras] {10.1093/mnras/stv1648}, \href
  {http://adsabs.harvard.edu/abs/2015MNRAS.453..671G} {453, 671}

\bibitem[\protect\citeauthoryear{{Gusakov}}{{Gusakov}}{2016}]{2016PhRvD..93f4033G}
{Gusakov} M.~E.,  2016, \mn@doi [\prd] {10.1103/PhysRevD.93.064033}, \href
  {http://adsabs.harvard.edu/abs/2016PhRvD..93f4033G} {93, 064033}

\bibitem[\protect\citeauthoryear{Gusakov \& Dommes}{Gusakov \&
  Dommes}{2016}]{2016arXiv160701629G}
Gusakov M.~E.,  Dommes V.~A.,  2016, \mn@doi [\prd]
  {10.1103/PhysRevD.94.083006}, 94, 083006

\bibitem[\protect\citeauthoryear{{Hall}}{{Hall}}{1960}]{1960AdPhy...9...89H}
{Hall} H.~E.,  1960, \mn@doi [Advances in Physics] {10.1080/00018736000101169},
  \href {http://adsabs.harvard.edu/abs/1960AdPhy...9...89H} {9, 89}

\bibitem[\protect\citeauthoryear{{Hall} \& {Vinen}}{{Hall} \&
  {Vinen}}{1956}]{1956RSPSA.238..215H}
{Hall} H.~E.,  {Vinen} W.~F.,  1956, \mn@doi [Proceedings of the Royal Society
  of London Series A] {10.1098/rspa.1956.0215}, \href
  {http://adsabs.harvard.edu/abs/1956RSPSA.238..215H} {238, 215}

\bibitem[\protect\citeauthoryear{{Harvey}, {Ruderman}  \& {Shaham}}{{Harvey}
  et~al.}{1986}]{hrs86}
{Harvey} J.~A.,  {Ruderman} M.~A.,   {Shaham} J.,  1986, \mn@doi [\prd]
  {10.1103/PhysRevD.33.2084}, \href
  {http://adsabs.harvard.edu/abs/1986PhRvD..33.2084H} {33, 2084}

\bibitem[\protect\citeauthoryear{{Jahan-Miri}}{{Jahan-Miri}}{2000}]{miri00}
{Jahan-Miri} M.,  2000, \mn@doi [\apj] {10.1086/308528}, \href
  {http://adsabs.harvard.edu/abs/2000ApJ...532..514J} {532, 514}

\bibitem[\protect\citeauthoryear{{Jahan-Miri}}{{Jahan-Miri}}{2002}]{miri02}
{Jahan-Miri} M.,  2002, \mn@doi [\prb] {10.1103/PhysRevB.65.184522}, \href
  {http://adsabs.harvard.edu/abs/2002PhRvB..65r4522J} {65, 184522}

\bibitem[\protect\citeauthoryear{{Jones}}{{Jones}}{1987}]{jones87}
{Jones} P.~B.,  1987, \mn@doi [\mnras] {10.1093/mnras/228.3.513}, \href
  {http://adsabs.harvard.edu/abs/1987MNRAS.228..513J} {228, 513}

\bibitem[\protect\citeauthoryear{{Jones}}{{Jones}}{2006}]{jones06}
{Jones} P.~B.,  2006, \mn@doi [\mnras] {10.1111/j.1365-2966.2005.09724.x},
  \href {http://adsabs.harvard.edu/abs/2006MNRAS.365..339J} {365, 339}

\bibitem[\protect\citeauthoryear{{Jones}}{{Jones}}{2009}]{jones09}
{Jones} P.~B.,  2009, \mn@doi [\mnras] {10.1111/j.1365-2966.2009.15016.x},
  \href {http://adsabs.harvard.edu/abs/2009MNRAS.397.1027J} {397, 1027}

\bibitem[\protect\citeauthoryear{Khalatnikov}{Khalatnikov}{2000}]{khalatnikov00}
Khalatnikov I.~M.,  2000, An introduction to the theory of superfluidity.
Westview Press, New York

\bibitem[\protect\citeauthoryear{{Konenkov} \& {Geppert}}{{Konenkov} \&
  {Geppert}}{2000}]{kg00}
{Konenkov} D.,  {Geppert} U.,  2000, \mn@doi [\mnras]
  {10.1046/j.1365-8711.2000.03188.x}, \href
  {http://adsabs.harvard.edu/abs/2000MNRAS.313...66K} {313, 66}

\bibitem[\protect\citeauthoryear{{Konenkov} \& {Geppert}}{{Konenkov} \&
  {Geppert}}{2001}]{kg01a}
{Konenkov} D.,  {Geppert} U.,  2001, \mn@doi [\mnras]
  {10.1046/j.1365-8711.2001.04469.x}, \href
  {http://adsabs.harvard.edu/abs/2001MNRAS.325..426K} {325, 426}

\bibitem[\protect\citeauthoryear{{Landau} \& {Lifshitz}}{{Landau} \&
  {Lifshitz}}{1980}]{ll80}
{Landau} L.~D.,  {Lifshitz} E.~M.,  1980, {Statistical physics. Pt.2}.
Pergamon Press, Oxford

\bibitem[\protect\citeauthoryear{{Lombardo} \& {Schulze}}{{Lombardo} \&
  {Schulze}}{2001}]{ls01}
{Lombardo} U.,  {Schulze} H.-J.,  2001, in {Blaschke} D.,  {Glendenning} N.~K.,
    {Sedrakian} A.,  eds,  Lecture Notes in Physics, Berlin Springer Verlag
  Vol. 578, Physics of Neutron Star Interiors. p.~30

\bibitem[\protect\citeauthoryear{{Mendell}}{{Mendell}}{1991}]{mendell91a}
{Mendell} G.,  1991, \mn@doi [\apj] {10.1086/170609}, \href
  {http://adsabs.harvard.edu/abs/1991ApJ...380..515M} {380, 515}

\bibitem[\protect\citeauthoryear{{Mendell} \& {Lindblom}}{{Mendell} \&
  {Lindblom}}{1991}]{1991AnPhy.205..110M}
{Mendell} G.,  {Lindblom} L.,  1991, \mn@doi [Annals of Physics]
  {10.1016/0003-4916(91)90239-5}, \href
  {http://adsabs.harvard.edu/abs/1991AnPhy.205..110M} {205, 110}

\bibitem[\protect\citeauthoryear{{Muslimov} \& {Tsygan}}{{Muslimov} \&
  {Tsygan}}{1985}]{1985Ap&SS.115...43M}
{Muslimov} A.~G.,  {Tsygan} A.~I.,  1985, \mn@doi [\apss] {10.1007/BF00653825},
  \href {http://adsabs.harvard.edu/abs/1985Ap%26SS.115...43M} {115, 43}

\bibitem[\protect\citeauthoryear{{Page}, {Lattimer}, {Prakash}  \&
  {Steiner}}{{Page} et~al.}{2013}]{plps13}
{Page} D.,  {Lattimer} J.~M.,  {Prakash} M.,   {Steiner} A.~W.,  2013,
  preprint, \href {http://adsabs.harvard.edu/abs/2013arXiv1302.6626P} {}
  (\mn@eprint {arXiv} {1302.6626})

\bibitem[\protect\citeauthoryear{{Rubinstein} \& {Pismen}}{{Rubinstein} \&
  {Pismen}}{1994}]{1994PhyD...78....1R}
{Rubinstein} B.~Y.,  {Pismen} L.~M.,  1994, \mn@doi [Physica D Nonlinear
  Phenomena] {10.1016/0167-2789(94)00119-7}, \href
  {http://adsabs.harvard.edu/abs/1994PhyD...78....1R} {78, 1}

\bibitem[\protect\citeauthoryear{{Sedrakian} \& {Sedrakian}}{{Sedrakian} \&
  {Sedrakian}}{1995}]{ss95}
{Sedrakian} A.~D.,  {Sedrakian} D.~M.,  1995, \mn@doi [\apj] {10.1086/175876},
  \href {http://adsabs.harvard.edu/abs/1995ApJ...447..305S} {447, 305}

\bibitem[\protect\citeauthoryear{{Sheehy} \& {Radzihovsky}}{{Sheehy} \&
  {Radzihovsky}}{2004}]{sr04}
{Sheehy} D.~E.,  {Radzihovsky} L.,  2004, \mn@doi [\pra]
  {10.1103/PhysRevA.70.063620}, \href
  {http://adsabs.harvard.edu/abs/2004PhRvA..70f3620S} {70, 063620}

\bibitem[\protect\citeauthoryear{{Sonin}}{{Sonin}}{1987}]{sonin87}
{Sonin} E.~B.,  1987, \mn@doi [Reviews of Modern Physics]
  {10.1103/RevModPhys.59.87}, \href
  {http://adsabs.harvard.edu/abs/1987RvMP...59...87S} {59, 87}

\bibitem[\protect\citeauthoryear{{Srinivasan}, {Bhattacharya}, {Muslimov}  \&
  {Tsygan}}{{Srinivasan} et~al.}{1990}]{sbmt90}
{Srinivasan} G.,  {Bhattacharya} D.,  {Muslimov} A.~G.,   {Tsygan} A.~J.,
  1990, Current Science, \href
  {http://adsabs.harvard.edu/abs/1990CSci...59...31S} {59, 31}

\makeatother
\end{thebibliography}
\bibliographystyle{mnras}

\end{document}